Quadrature skyrmions in two-dimensionally arrayed parametric resonators

Hiroshi Yamaguchi, Daiki Hatanaka, and Motoki Asano
NTT Basic Research Laboratories, Atsugi, Kanagawa 243-0198, Japan

Skyrmions are topological solitons in two-dimensional systems and have been observed in various physical systems. Generating and controlling skyrmions in artificial resonator arrays lead to novel acoustic, photonic, and electric devices, but it is a challenge to implement a vector variable with the chiral exchange interaction. Here, we propose to use quadrature variables, where their parametric coupling enables skyrmions to be stabilized. A finite-element simulation indicates that a acoustic skyrmion would exist in a realistic structure consisting of a piezoelectric membrane array.

Topological solitons are fundamental excitations observed in various nonlinear systems [1]. The soliton is the phase boundary between two identical but topologically distinguished domains and can be created by spontaneous breaking of translational symmetry [2,3]. One of the most important two-dimensional (2D) solitons is the skyrmion, which has recently attracted much attention especially in the field of ferromagnetic systems [4-10]. A skyrmion has a chiral spin texture and can be induced by an antisymmetric exchange interaction together with a symmetric exchange coupling [11-13]. Skyrmions are stable because of their topological nature and a scheme to control their motion has been proposed [14,15] which may lead to applications of skyrmions in memory [16,17], logic [18], and microwave devices [19-21]. Skyrmions have been shown to appear in numerous physical systems, including quantum Hall systems [22-24] and Bose-Einstein condensate [25-27] and are considered universal phenomena.

The concept of topological insulators has been recently utilized in artificial 2D lattice systems of photonics [28-30], microwave [31], and acoustics [32-34]. The generation and control of skyrmions in these artificial systems are expected to similarly lead to new device and material technologies. However, there have been few reports on the skyrmions generated by spontaneous breaking of translational symmetry in artificial structures. One of the major challenges is how to construct the 2D array of three-component vector variables, with both symmetric and antisymmetric exchange interactions. Various efforts have been made using velocity fields [35] and hybrid displacements [36] in phononics, and also evanescence fields in photonics [37-39], but in the majority of the studies, skyrmion-like textures were created not by spontaneous symmetry breaking but by interference between external drive signals. Therefore, in those studies the position of each skyrmion is externally specified and no motion can be induced without changing the external drive. Thus, finding artificial lattice systems with a chiral exchange interaction to induce the spontaneous breaking remains an important research target in the field of skyrmion physics.

In this letter, we propose an alternative approach that utilizes quadrature variables [40], instead of three real-space vector components, in a doubly degenerate 2D harmonic resonator array. This approach has three significant features. First, the skyrmion texture can be generated with the array of only two-component variables. Second, the periodic temporal perturbation, which so far has been discussed mainly in terms of linear Floquet systems [41-43], is also a useful scheme in nonlinear dynamics. Finally, as in the case of topological insulators, the stability of the texture induced by the topological properties can be utilized to develop novel devices and materials technologies. The approach can be applied to various artificial 2D-arrayed resonators, such as microwave circuits and photonic and acoustic metamaterials, holding out the prospect of extensive applications of skyrmion physics to device and material technologies.

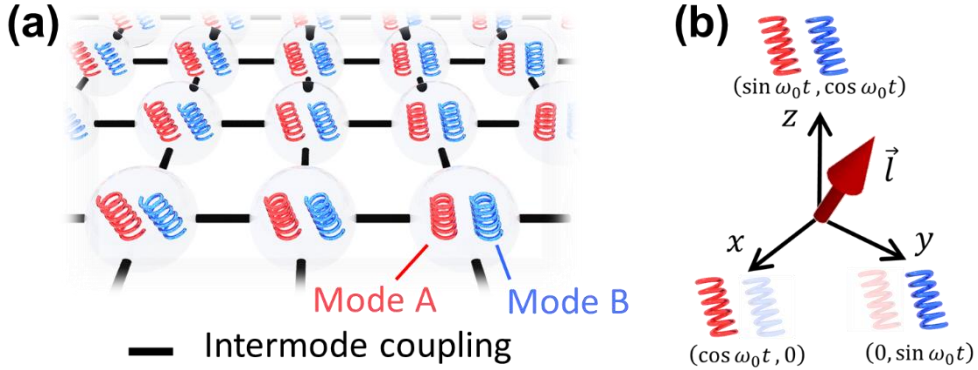

FIG. 1. Conceptual drawing of (a) a doubly degenerate resonator array and (b) a quadrature moment. The $x$-, $y$-, and $z$-aligned $\vec{l}$ quadrature moments correspond to the oscillation states of $(q_A, q_B) = (a_0 \cos \omega_0 t, 0)$, $(0, a_0 \sin \omega_0 t)$, and $(a_0 \sin \omega_0 t, a_0 \cos \omega_0 t)/\sqrt{2}$, respectively.

We consider a square array with doubly degenerate resonance modes $A$ and $B$ (Fig. 1(a)). The mode variables $q_k(t)$ at a particular site on the lattice is expressed as
$$q_k(t) = c_k(t) \cos \omega_0 t + s_k(t) \sin \omega_0 t, \quad (k = A, B), \tag{1}$$
under the rotating frame approximation, where the cosine and sine oscillation amplitudes, $c_k$ and $s_k$, are called quadratures. $\omega_0$ is the angular frequency of the two modes. We introduce a parametric excitation at twice the frequency, $2\omega_0$, and consider the case that the excitation changes sign between the two orthogonal modes. This is the essential assumption under which to assign stable parametric oscillation states to z-polarized vector states. The physical implementation will be described later. The laboratory-frame Hamiltonian $H_0$ is given by

$$H_0 = \frac{p_A^2 + p_B^2}{2} + \frac{\omega_0^2}{2}(q_A^2 + q_B^2) + \omega_0^2 \Gamma \cos 2\omega_0 t \, (q_A^2 - q_B^2) + \frac{\omega_0^2 \alpha}{4}(q_A^2 + q_B^2)^2. \quad (2)$$

Here, $p_k = \dot{q}_k$ is the canonically conjugate momentum of $q_k$. We have assumed isotropic cubic nonlinearity with a strength of $\alpha$ and $\Gamma$ is the parametric excitation intensity. We also assume identical effective masses for both modes and use the unit to make it unity for simplicity. The new Hamiltonian $h_0$ is expressed as

$$h_0 \sim \frac{\omega_0^2 \Gamma}{4}[(c_A^2 - s_A^2) - (c_B^2 - s_B^2)] + \frac{3\omega_0^2 \alpha}{32}[(c_A^2 + s_A^2) + (c_B^2 + s_B^2)]^2$$
$$- \frac{\omega_0^2 \alpha}{8}(c_A s_B - s_A c_B)^2 \quad (3)$$

and the equation of motion is given by $\dot{c}_k = \omega_0^{-1} \partial h_0 / \partial s_k$ and $\dot{s}_k = -\omega_0^{-1} \partial h_0 / \partial c_k$ (See the Supplemental Materials (S.M.)). The steady-state solution, $\dot{c}_k = \dot{s}_k = 0$, corresponds to a local minimum (or maximum) of $h_0$ [40,44] and consists of four oscillation states, $L_\pm$ and $R_\pm$, given by $(c_A, s_A, c_B, s_B) = \sqrt{\Gamma/\alpha}(0, \pm 1, \pm 1, 0)$ and $\sqrt{\Gamma/\alpha}(0, \pm 1, \mp 1, 0)$, i.e.

$$\begin{array}{ll} L\pm: & q_A(t) = \pm a_0 \sin \omega_0 t, \quad q_B(t) = \pm a_0 \cos \omega_0 t \\ R\pm: & q_A(t) = \pm a_0 \sin \omega_0 t, \quad q_B(t) = \mp a_0 \cos \omega_0 t \end{array} \quad (a_0 = \sqrt{\Gamma/\alpha}). \quad (4)$$

As will be shown later, $L \pm$ and $R \pm$ are clockwise and anti-clockwise circularly polarized oscillation states, where the sign corresponds to the parametric oscillation phases. This Hamiltonian picture provides a good approximation even for a realistic case that has a finite damping, because it assumes that the parametric actuation $\Gamma$ is sufficiently larger than the oscillation threshold determined by the damping. Because of the system nonlinearity and the $\pi$ phase difference in parametric excitation between modes A and B, the two modes are equally mixed with a $\pi/2$ phase difference and form stable circularly polarized oscillations.

Next, we define a three-component quadrature moment using four variables, $c_A, s_A, c_B, s_B$. We employ two definitions in which the moment is linear with respect to the quadratures,

$$\vec{l} = (l_x, l_y, l_z) \equiv \left(c_A, s_B, \frac{s_A + c_B}{\sqrt{2}}\right), \quad \vec{r} = (r_x, r_y, r_z) \equiv \left(c_A, s_B, \frac{s_A - c_B}{\sqrt{2}}\right). \quad (5)$$

These definitions do not satisfy the Poisson-bracket algebra of angular momentum but are ideal for our purpose of creating skyrmions. We can easily confirm that the four parametric oscillation states $L \pm$ and $R \pm$ correspond to z-polarized quadrature moments (See Fig. 1(b) and S.M.). It should be noted that $h_0$ is invariant under the replacement, $(c_A \quad s_B) \to (s_B \quad -c_A)$, which corresponds to a $\pi/2$ rotation in the $xy$ plane as confirmed by the definition (5).

Next, as a preliminary step before discussing the detailed device structures, we mathematically construct a Dzyaloshinskii–Moriya (DM) interaction for the quadrature moments as well as the symmetric exchange coupling and numerically confirm that it leads to a skyrmion texture under parametric excitation. The symmetric exchange coupling between two nearest-neighbor sites is given by the isotropic interaction, $h_{EX} = -g_s(q_{1A}q_{2A} + q_{1B}q_{2B})$ with $g_s > 0$, where the indexes 1 and 2 specify adjacent site positions. The rotating frame approximation leads to

$$h_{EX} \sim -\frac{g_s}{2}(c_{1A}c_{2A} + s_{1A}s_{2A} + c_{1B}c_{2B} + s_{1B}s_{2B}). \tag{6}$$

Then, we introduce a chiral exchange interaction to create a skyrmion texture. The DM interaction between two nearest-neighbor magnetic moments $\vec{m}_1$ and $\vec{m}_2$ is expressed as $h_{DM} = g_{DM}\vec{n}_{12} \cdot (\vec{m}_1 \times \vec{m}_2)$ [11,12]. Here, $\vec{n}_{12}$ is the DM vector and it can be chosen as, for example, a unit vector directed from site 1 to site 2 for generating a Bloch-type skyrmion [45]. For a resonator pair parallel to the x-axis, $\vec{n}_{12} = e_x \equiv (1,0,0)$ and $h_{DM-x} = g_{DM}(m_1^y m_2^z - m_1^z m_2^y)$. Similarly, for the y-axis, $\vec{n}_{12} = e_y \equiv (0,1,0)$ and $h_{DM-y} = g_{DM}(m_1^z m_2^x - m_1^x m_2^z)$. If we replace $\vec{m}$ by $\vec{l}$ or $\vec{r}$, we obtain the DM interaction from Eq. (5) as follows:

$$h_{DM-x}^{(L/R)} = \frac{g_{DM}}{\sqrt{2}}[s_{1B}s_{2A} - s_{1A}s_{2B} \pm (s_{1B}c_{2B} - c_{1B}s_{2B})],$$

$$h_{DM-y}^{(L/R)} = \frac{g_{DM}}{\sqrt{2}}[s_{1A}c_{2A} - c_{1A}s_{2A} \pm (c_{1B}c_{2A} - c_{1A}c_{2B})]. \tag{7}$$

Here, the plus (minus) sign corresponds to the $L$ ($R$) mode. The terms mixing sine and cosine quadratures need a phase-shifted coupling. We can eliminate these terms by making a linear combination,

$$h_{DM-x} = \frac{h_{DM-x}^{(L)} + h_{DM-x}^{(R)}}{\sqrt{2}} = g_{DM}[s_{1B}s_{2A} - s_{1A}s_{2B}],$$

$$h_{DM-y} = \frac{h_{DM-y}^{(L)} - h_{DM-y}^{(R)}}{\sqrt{2}} = g_{DM}[c_{1B}c_{2A} - c_{1A}c_{2B}]. \tag{8}$$

For a $\pi/2$ rotation of the quadrature moments, i.e. $(c_A \quad s_B) \to (s_B \quad -c_A)$, this interaction sustains symmetry for $L \pm$ states and antisymmetry for $R \pm$. This difference induces different skyrmion textures, Bloch skyrmions for $L \pm$ but antiskyrmions for $R \pm$, as shown below. Although this DM interaction seems artificial, we will later find that it can be implemented in a realistic mechanical resonator array.

We then performed a numerical calculation to find a stable texture of quadrature moments. Hereafter, we will employ the unit $\omega_0 = 1$ for simplicity. The full interaction Hamiltonian is given by

$$h_{int} = -\frac{g_s}{2} \sum_{i,j,k=A,B} \left( c_{(i,j)k} c_{(i+1,j)k} + c_{(i,j)k} c_{(i,j+1)k} + s_{(i,j)k} s_{(i+1,j)k} + s_{(i,j)k} s_{(i,j+1)k} \right)$$
$$+ g_{DM} \sum_{i,j} \left( s_{(i,j)B} s_{(i+1,j)A} - s_{(i,j)A} s_{(i+1,j)B} + c_{(i,j)B} c_{(i,j+1)A} - c_{(i,j)A} c_{(i,j+1)B} \right) \quad (9)$$

Here, the suffix $(i,j)$ denotes the site position. A stable solution that minimizes the total Hamiltonian was calculated starting from the initial texture, in which an $L+$ domain of radius $n_r$ is surrounded by an $L-$ domain. The time evolution was calculated using the Runge-Kutta method after applying a small perturbation to the obtained minimum-energy solution. This test confirms the local stability of obtained solutions (S.M.). The calculated textures using $64 \times 64$ unit cells for various $g_{DM}$ are shown in Fig. 2(a) in the form of the 2D distribution of $\vec{l}_{(i,j)}$.

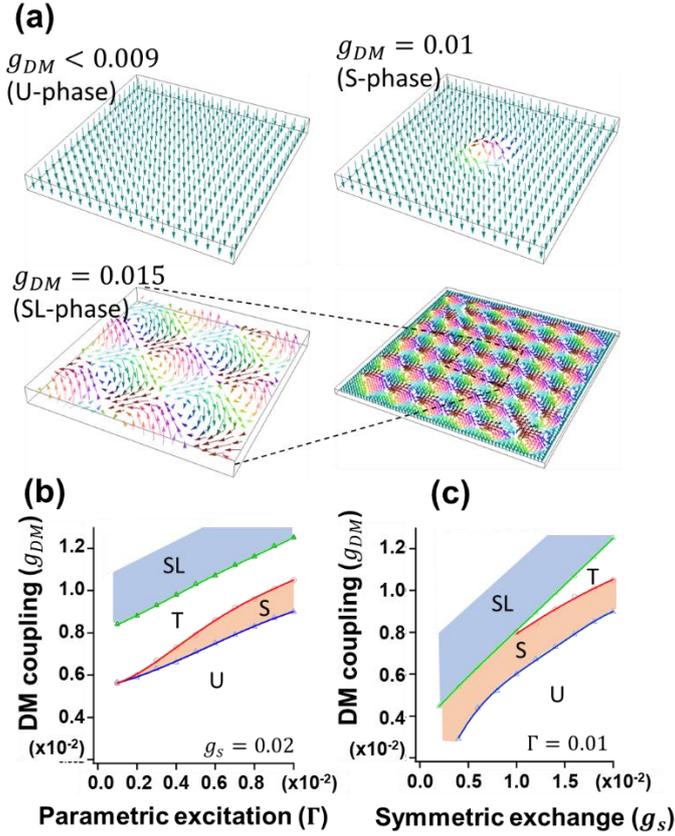

FIG. 2. (a) Calculated spatial distribution of $\vec{l}_{(i,j)}$ for various $g_{DM}$ with $g_s = 0.02$, $\Gamma = 0.01$, and $n_r = 2$. The color corresponds to the direction and amplitude of quadrature moments. RGB intensities correspond to $l_z$, $l_x$, and $l_y$ components, respectively. (b)

and (c) Phase diagrams of unstable (U), skyrmion (S), transition (T), and skyrmion lattice (SL) regions for various $g_{DM}, \Gamma$, and $g_S$.

When $g_{DM}$ is smaller than the onset value for skyrmion formation ($g_{DM} < 0.009$), the skyrmion texture is unstable (U-phase) and all of the moments are directed downward, i.e. $\vec{l}_{(i,j)} = (0,0,-\sqrt{2}a_0)$ for all $(i,j)$. Above the onset, a single skyrmion texture forms (S-phase) whose structure is nearly independent of $g_{DM}$. The calculated time evolution shows that the texture is metastable because a perturbation of 5% of the oscillation amplitude destabilizes it, whereas it remains stable against the smaller perturbation. Above the upper bound ($g_{DM} \geq 0.0105$), the transition to an extended texture (T-phase) emerges, where the skyrmion deconfines and occupies the whole lattice in an extremely anisotropic shape, and finally the texture of a skyrmion lattice (SL-phase) forms at $g_{DM} \geq 0.0125$. This texture is independent of the initial guess in the calculation in contrast to the S-phase, so that the lattice remains stable, at least within our numerical calculations.

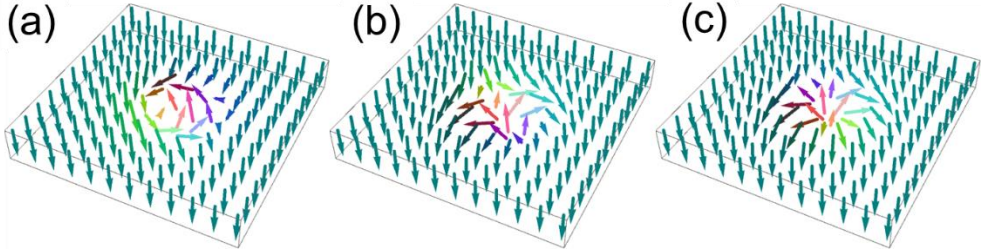

FIG. 3. Birds-eye views of calculated textures for (a) Bloch skyrmion, (b) antiskyrmion, and (c) Neel skyrmion, plotting $\vec{l}_{(i,j)}$ for (a), and $\vec{r}_{(i,j)}$ for (b) and (c). The DM vector is $\vec{n}_{12} = e_x$ for the x-axis and $e_y$ for the y-axis in (a) and (b), whereas it is $e_y$ for the x-axis and $-e_x$ for the y-axis in (c). For all plots, $g_s = 0.02$, $\Gamma = 0.01$, $g_{DM} = 0.095$ and $n_r = 2$.

We examined the topological features of the obtained textures by calculating the skyrmion number. Instead of the generally used expression for a continuous system, $S = \int \vec{n} \cdot (\partial_x \vec{n} \times \partial_y \vec{n}) dxdy / 4\pi$, summing up the solid angles formed by all sets of three adjacent pseudo moments makes the skyrmion number unity as expected for a topological soliton (S.M.). Figure 2(b) and (c) show the phase diagrams of the texture as a function of $\Gamma$, $g_s$, and $g_{DM}$. The S-U boundary in Fig. 2(c) suggests that the parameter $g_{DM}^2/g_s$ governs the transition [45]. We can see that the skyrmion and its lattice textures occur over a wide parameter range. Similar calculations using $\vec{r}$ generated an antiskyrmion texture (Fig. 3 (b)). We also generated a Neel-type skyrmion by using the different DM vector, $\vec{n}_{12} = e_y$ for the x-axis and $-e_x$ for the y-axis (Fig. 3(c)) [45], indicating that various skyrmion textures are possible.

Next, let us discuss a realistic device model implementing the DM interaction. Although our concept can be applied to various physical systems, we here envision an implementation in a nanomechanical resonator array consisting of coupled piezoelectric circular membranes [47]. The unit structure, shown in Fig. 4(a), has doubly-degenerate modes, A and B (Fig. 4(b)), and the circular edge is fixed. Because of its $d_{31}$ piezoelectric component [48,49], the applied voltage induces a different sign of tension between two orthogonal directions, $[110]$ and $[1\bar{1}0]$ (Fig. 4(a) and S.M.). This anisotropy changes the sign of parametric excitation required to derive the third term in Eq. (2) and enables four circularly polarized parametric oscillation states to be excited (Eq. (4) and Fig. 4(c)).

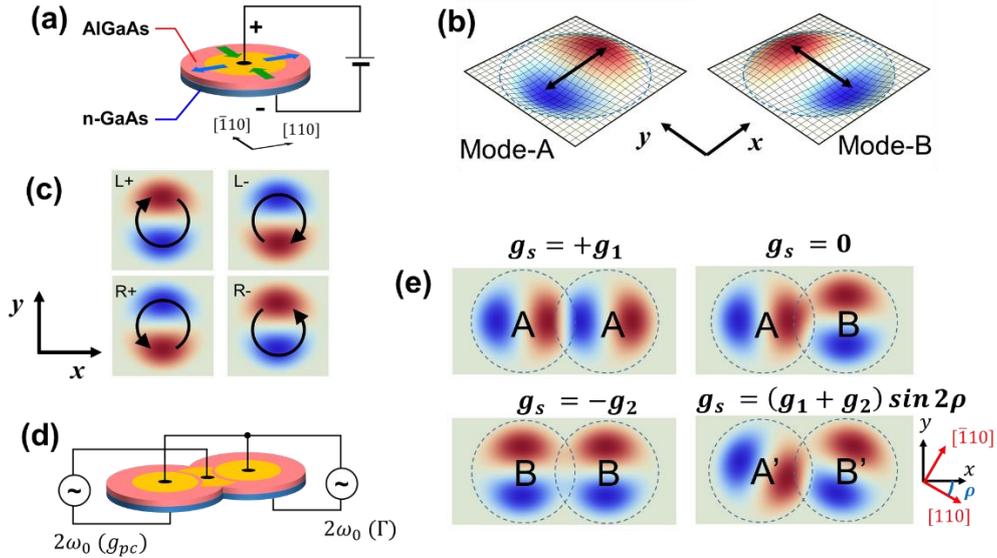

FIG. 4. (a) Schematic drawing of circular electromechanical membrane resonator and (b) two vibration modes, A and B. (c) Four parametric oscillation states. L/R indicates the circular polarization and $+/-$ indicates the oscillation phase. (d) Schematic drawing of the device configuration. An alternating voltage at $2\omega_0$ is applied to the center electrode for parametric excitation and to the junction electrode for parametric coupling. (e) Schematic illustration of the coupling constants. Two different constants, $g_1$ and $g_2$, are introduced and the coupling between the A and B modes is established by misaligning the piezoelectric and membrane axis.

The symmetric exchange coupling, Eq. (6), is mediated by the elastic coupling through the overlap (Fig. 4(d)), where we need to consider two coupling constants, $g_1$ and $g_2$ (Fig. 4(e)). A comparison of (8) with (6) indicates that a $AB$ mixing interaction is required to induce the quadrature DM interaction. Accordingly, we employed a misalignment between the piezoelectric and membrane-lattice axis (Fig. 4(e)). After redefining the even-site resonator amplitude with reversed sign, the mixing term is given by

$$[h_{EX-\rho}]_{AB} \sim -\frac{g_S}{2}\sin 2\rho\, (c_{1B}c_{2A} + s_{1B}s_{2A} - c_{1A}c_{2B} - s_{1A}s_{2B}). \tag{10}$$

Here, $\rho$ is the misalignment angle and $g_S = (g_1 + g_2)/2$ (see full expression in S.M.). Although the $AB$ mixing is introduced, this interaction is identical between the x- and y-axis and unable to introduce the chiral interaction. This is reasonable because both $c_k$'s and $s_k$'s are coupled in the same way through $q_k$'s. To induce chirality, we need to introduce an interaction that independently couples $c_k$'s and $s_k$'s. This can be done by using the parametric coupling proposed for another topological system [30]. We here show that combining this parametric coupling together with the parametric excitation induces a stable skyrmion texture.

A modulation, $g_S(t) = g_{S0} + g_{pc}\cos(2\omega_0 t)$, leads to the following $AB$ mixing exchange interaction,

$$[h_{pc}]_{AB} = -\frac{g_{pc}}{4}\sin 2\rho\, (c_{1B}c_{2A} - s_{1B}s_{2A} - c_{1A}c_{2B} + s_{1A}s_{2B}). \tag{11}$$

Here, the signs of $s_k$ are opposite to those in (10) so that applying a different signed $2\omega_0$ modulation between the x- and y-axis coupling creates a chiral interaction. This can easily be done in our device configuration by applying opposite signed alternating voltages. The total $AB$ mixing part becomes

$$[h_{EX-\rho} + h_{pc}]_{AB} = -\sin 2\rho\left[\frac{g_+}{2}(c_{1B}c_{2A} - c_{1A}c_{2B}) + \frac{g_-}{2}(s_{1B}s_{2A} - s_{1A}s_{2B})\right] \tag{12}$$

Here, $g_+ = g_{S0} + \frac{g_{pc}}{2}$, $g_- = g_{S0} - \frac{g_{pc}}{2}$. In the case of $g_{pc} = -2g_{S0}$ for the *x*-axis and $+2g_{S0}$ for the *y*-axis, we obtain

$$[h_{EX-\rho} + h_{pc}]_{AB:x} = -g_{S0}\sin 2\rho\,(s_{1B}s_{2A} - s_{1A}s_{2B})$$
$$[h_{EX-\rho} + h_{pc}]_{AB:y} = -g_{S0}\sin 2\rho\,(c_{1B}c_{2A} - c_{1A}c_{2B}) \tag{13}$$

These equations are identical to the DM interaction, Eq. (8). Therefore, the combined effects of the misaligned piezoelectric axis and $2\omega_0$ modulation of the symmetric exchange coupling can induce the chiral interaction required to form a skyrmion. We can also generate a Neel-type DM interaction by changing the sign of the parametric coupling (S.M.), so that the kind of creating skyrmion can be externally controlled.

We performed numerical calculations to verify the above findings. First, we performed a simulation using the finite element method (FEM) using COMSOL Multiphysics® to numerically determine the coupling constants. Figure 5(a) shows the calculated coupling constant between two adjacent membrane resonators as a function of gate voltage on the overlap area. For an offset bias voltage of 20 V, we can successfully obtain a Neel-type texture as shown in Fig. 5(b).

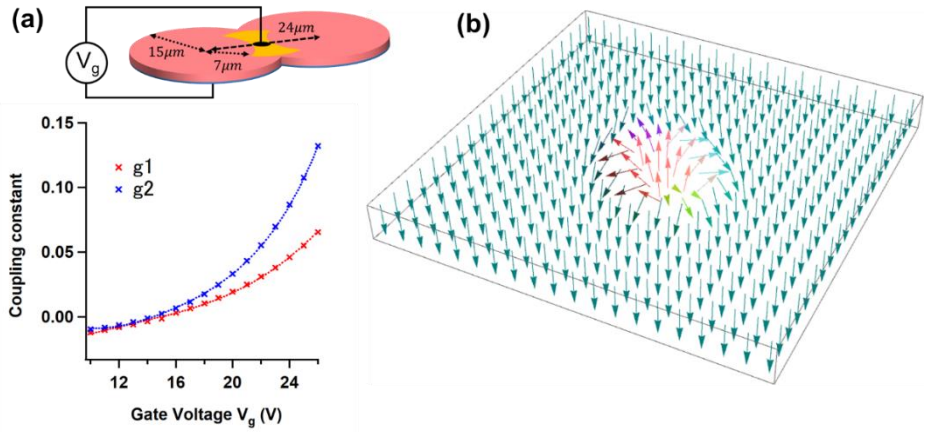

FIG. 5. (a) Calculated static coupling constants as a function of gate voltage using FEM for the geometry indicated at the top. The resonator thickness is 100 nm and the elastic and piezoelectric parameters of GaAs are used in the calculation. (b) The texture of Neel skyrmion in terms of $\vec{r}_{(i,j)}$ calculated from the parameters given by FEM and using a voltage modulation of $4.0\ V_{rms}$ and $\rho = 20°$.

Finally, let us discuss the robustness of skyrmions against external fluctuations. Fabrication inaccuracies affecting the resonance frequency are some of the most common causes of fluctuations. We used a model employing the ideal DM interaction with Eq. (9) and the steady-state textures for various resonance frequency fluctuation $\sigma = \sqrt{\overline{(\omega_{0(i,j)} - \omega_0)^2}}/\omega_0$ are calculated (S.M.). When $\sigma \leq 0.02$, the skyrmion texture remains metastable against the frequency fluctuation. For a larger $\sigma$, skyrmion-shaped excitations were randomly generated, indicating instability. When the fluctuation is larger than $g_s$, ferromagnetic ordering is not maintained. In addition, the degeneracy lifting between A and B modes should be within the resonance linewidth to maintain circularly polarized modes. Nonetheless, since these fabrication tolerances are easily obtained [50], it should be possible to demonstrate acoustic skyrmions in actual devices.

In conclusion, we proposed an approach to generating various skyrmions that utilizes four quadrature variables in 2D parametric resonator arrays. This is the first numerical demonstration of skyrmions in systems formed by only two dynamical variables through temporal periodic perturbation. The concept can be applied to other kinds of parametric resonator, including photonic and microwave resonator arrays.

The authors would like to thank Hajime Okamoto for his continuous encouragements. This work was partially supported by JSPS KAKENHI Grant Number JP21H05020 and JP23H05463.

# Supplemental materials

## 1. Derivation of parametric oscillation states

We start from the single resonator Hamiltonian (2),

$$H_0 = \frac{p_A^2 + p_B^2}{2} + \frac{\omega_0^2}{2}(q_A^2 + q_B^2) + \omega_0^2 \Gamma \cos 2\omega_0 t \, (q_A^2 - q_B^2)$$
$$+ \frac{\omega_0^2 \alpha}{4}(q_A^2 + q_B^2)^2.$$

Using the generator $F$ of the time-dependent canonical transformation for a rotating frame,

$$F(q_A, Q_A, q_B, Q_B, t) = \sum_{k=A,B} \left( \frac{\omega_0 q_k^2}{2 \tan \omega_0 t} - \frac{\sqrt{\omega_0} q_k Q_k}{\sin \omega_0 t} + \frac{Q_k^2}{2 \tan \omega_0 t} \right)$$

$$p_k = \frac{\partial F}{\partial q_k}, \qquad P_k = -\frac{\partial F}{\partial Q_k}, \qquad h_0 = H_0 + \frac{\partial F}{\partial t},$$

we obtain

$$q_k = \omega_0^{-1/2}(P_k \sin \omega_0 t + Q_k \cos \omega_0 t),$$
$$p_k = \omega_0^{1/2}(P_k \cos \omega_0 t - Q_k \sin \omega_0 t),$$
$$h_0 = \frac{3\alpha}{32}(P_A^2 + Q_A^2 + P_B^2 + Q_B^2)^2 - \frac{\alpha}{8}(Q_A P_B - P_A Q_B)^2.$$

Here, we neglect the rapidly oscillating terms. The canonical equations of motion become

$$\dot{Q}_k = \frac{\partial h_0}{\partial P_k}, \qquad \dot{P}_k = -\frac{\partial h_0}{\partial Q_k}$$

We can obtain the equation of motion by using $c_k = \omega_0^{-1/2} Q_k$ and $s_k = \omega_0^{-1/2} P_k$. To find the steady-state solution, we define $c$, $s$, $\theta_c$ and $\theta_s$ as

$$c_A = c \cos \theta_c, \, s_B = c \sin \theta_c$$
$$s_A = s \cos \theta_s, \, c_B = s \sin \theta_s$$

Then,

$$h_0 = \omega_0^2 \left[ \frac{\Gamma}{4}(c^2 - s^2) + \frac{3\alpha}{32}(c^2 + s^2)^2 - \frac{\alpha}{32}(c^2 \sin 2\theta_c - s^2 \sin 2\theta_s)^2 \right]$$

We assume that $\Gamma > 0$ and $\alpha > 0$; then $h_0 \to +\infty$ for $c, s \to \pm\infty$. Therefore, a stable solution is found at the minimum of $h_0$. For given $c$ and $s$, the last term becomes smallest when

$$\sin 2\theta_c = \pm 1, \quad \sin 2\theta_s = 0 \quad (c^2 \geq s^2)$$
$$\sin 2\theta_c = 0, \quad \sin 2\theta_s = \pm 1 \quad (c^2 \leq s^2)$$

Then,

$$h_0 = \omega_0{}^2 \left[\frac{\Gamma}{4}(c^2 - s^2) + \frac{3\alpha}{32}(c^2 + s^2)^2 - \frac{\alpha}{32}c^4\right] \quad (c^2 \geq s^2)$$

$$h_0 = \omega_0{}^2 \left[\frac{\Gamma}{4}(c^2 - s^2) + \frac{3\alpha}{32}(c^2 + s^2)^2 - \frac{\alpha}{32}s^4\right] \quad (c^2 \leq s^2)$$

When $c^2 \geq s^2$, the coefficients of $c^4$ and $c^2$ are always positive so that the energy has a minimum value when $c = 0$ and $s = 0$. Therefore, $c^2 < s^2$. Then, $c = 0$ and

$$h_0 = \omega_0{}^2 \left[\frac{\alpha}{16}s^4 - \frac{\Gamma}{4}s^2\right] = \frac{\alpha \omega_0{}^2}{16}\left[\left(s^2 - \frac{2\Gamma}{\alpha}\right)^2 - \frac{4\Gamma^2}{\alpha^2}\right].$$

This expression reaches a minimum at

$$s = \pm\sqrt{2}a_0, \quad a_0 = \sqrt{\frac{\Gamma}{\alpha}}.$$

On the other hand, from $\sin 2\theta_s = \pm 1$, we have $\theta_s = \frac{\pi}{4}, \frac{3\pi}{4}, \frac{5\pi}{4}, \frac{7\pi}{4}$. Thus, the four solutions are

$$c_A = 0, \quad s_B = 0, \quad s_A = s\cos\theta_s = \pm a_0, \quad c_B = s\sin\theta_s = \pm a_0.$$

Here, the sign $\pm$ can be chosen independently. The above correspond to the solutions of eq. (4). The expression using the quadrature moments $\vec{l}$ and $\vec{r}$ is shown in Table S1, indicating that $L\pm$ and $R\pm$ correspond to z-polarized quadrature moments.

|   | $(l_x, l_y, l_z)$ | $(r_x, r_y, r_z)$ |
|---|---|---|
| $L\pm$ | $(0,0,\pm\sqrt{2}a_0)$ | $(0,0,0)$ |
| $R\pm$ | $(0,0,0)$ | $(0,0,\pm\sqrt{2}a_0)$ |

Table S1: Quadrature moments for four parametric oscillation states, $L\pm$ and $R\pm$.

## 2. Stability test of skyrmion textures

We perform a two-step calculation to investigate the stability of the skyrmion solution. First, a solution providing the local minimum of $h_0$ is calculated. Then, small perturbations are added to the quadratures,

$$c_{(i,j)k} \to c_{(i,j)k} + \Delta c_{(i,j)k}$$
$$s_{(i,j)k} \to s_{(i,j)k} + \Delta s_{(i,j)k}$$

at $t=0$, and their time evolution is calculated, where the equations of motion are modified to include the effect of weak damping,

$$\dot{c}_k = \omega_0^{-1}\frac{\partial h_0}{\partial s_k} - \frac{\omega_0 Q}{2}c_k, \qquad \dot{s}_k = -\omega_0^{-1}\frac{\partial h_0}{\partial c_k} - \frac{\omega_0 Q}{2}s_k,$$

where $Q$ is the quality factor of the resonator (assumed to be $10^4$). Figure S1 plots the time evolution of the skyrmion area, i.e. the average number of positively z-polarized quadrature moments for different amounts of perturbation $\varepsilon = \sqrt{\overline{(\Delta c_{(i,j)k})^2}} = \sqrt{\overline{(\Delta s_{(i,j)k})^2}}$. The texture is maintained for $\varepsilon \leq 6\%$, but it disappears for $\varepsilon \geq 7\%$. This result confirms that the skyrmion is locally stable but a large perturbation breaks the texture, causing the system to fall into the lowest-energy single ferromagnetic domain. Therefore, the isolated skyrmion is not energetically lowest but metastable, corresponding to the locally minimum pseudo-energy state. We compared the textures using various initial guess, changing the position, radius, and shape of oppositely polarized domains and no significant change in phase diagram was observed.

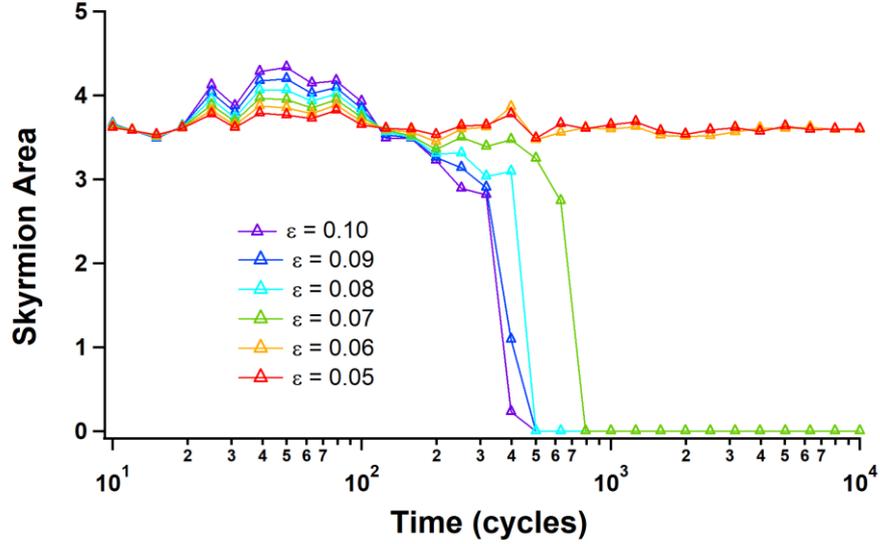

Fig. S1. Time evolution of skyrmion area with a typical texture. The unit of time is the oscillation cycle, i.e. $\omega_0 t/2\pi$.

Fig. S2 shows the displacement distribution of the membrane resonator array calculated from the steady-state quadrature moments. A movie showing the time evolution is in a separate file.

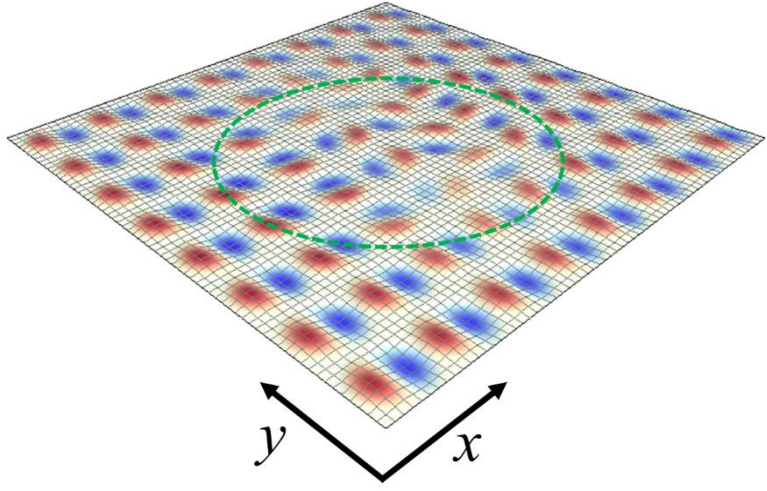

Fig. S2. Snapshot of displacement distribution of membrane resonator array calculated from the steady-state quadrature moments of the Bloch skyrmion shown in Fig. 3(a). The green dashed line roughly indicates the boundary of the skyrmion.

3. Calculation of skyrmion number

The skyrmion number of a continuous system is calculated using the formula,

$$S = \frac{\int \vec{n} \cdot (\partial_x \vec{n} \times \partial_y \vec{n}) dx dy}{4\pi}$$

Here, $\vec{n}$ is the unit vector of the moment. The discrete version is given by

$$S = \frac{\sum_{i,j} \vec{\eta}_{(i,j)} \cdot (\Delta_x \vec{\eta}_{(i,j)} \times \Delta_y \vec{\eta}_{(i,j)})}{4\pi} = \frac{\sum_{i,j} \vec{\eta}_{(i,j)} \cdot (\vec{\eta}_{(i+1,j)} \times \vec{\eta}_{(i,j+1)})}{4\pi} \quad (S1)$$

Here, $\vec{\eta}_{(i,j)} \equiv \vec{l}_{(i,j)}/|\vec{l}_{(i,j)}|$ (in the case of $\vec{l}$) is the unit vector of the quadrature moment and $\Delta_x \vec{\eta}_{(i,j)} = \vec{\eta}_{(i+1,j)} - \vec{\eta}_{(i,j)}$ and $\Delta_y \vec{\eta}_{(i,j)} = \vec{\eta}_{(i,j+1)} - \vec{\eta}_{(i,j)}$ are nearest-neighbor differences. We obtained $C = 0.65$ for an isolated skyrmion ($g_{DM} = 0.01$ in Fig. 2(a)). This is smaller than the ideal value of unity because of the discreteness of the lattice. By using L'Huilier's formula (Ref. [S1]), which determines the solid angle $\Omega$ (Fig. S3.) from three unit vectors,

$$\Omega = 4\eta \tan^{-1} \sqrt{\tan\frac{\theta_s}{2} \tan\frac{\theta_s - \theta_A}{2} \tan\frac{\theta_s - \theta_B}{2} \tan\frac{\theta_s - \theta_C}{2}}$$

$$\theta_s = \frac{\theta_A + \theta_B + \theta_C}{2}$$

we can calculate the total solid angle covered by the quadrature moments for all the sets of

$$(\cos\theta_A, \cos\theta_B, \cos\theta_C) = (\vec{\eta}_{(i+1,j)} \cdot \vec{\eta}_{(i,j+1)}, \vec{\eta}_{(i,j)} \cdot \vec{\eta}_{(i+1,j)}, \vec{\eta}_{(i,j)} \cdot \vec{\eta}_{(i,j+1)}),$$
$$\eta = \text{sgn}[(\vec{\eta}_{(i+1,j)} \times \vec{\eta}_{(i,j+1)}) \cdot \vec{\eta}_{(i,j)}],$$

and

$$(\cos\theta_A, \cos\theta_B, \cos\theta_C) = (\vec{\eta}_{(i-1,j)} \cdot \vec{\eta}_{(i,j-1)}, \vec{\eta}_{(i,j)} \cdot \vec{\eta}_{(i-1,j)}, \vec{\eta}_{(i,j)} \cdot \vec{\eta}_{(i,j-1)}),$$
$$\eta = \text{sgn}[(\vec{\eta}_{(i-1,j)} \times \vec{\eta}_{(i,j-1)}) \cdot \vec{\eta}_{(i,j)}].$$

The sum of $\Omega$ for the obtained isolated skyrmion solution becomes $4\pi$, showing that the pseudo moment covers the surface of a unit sphere.

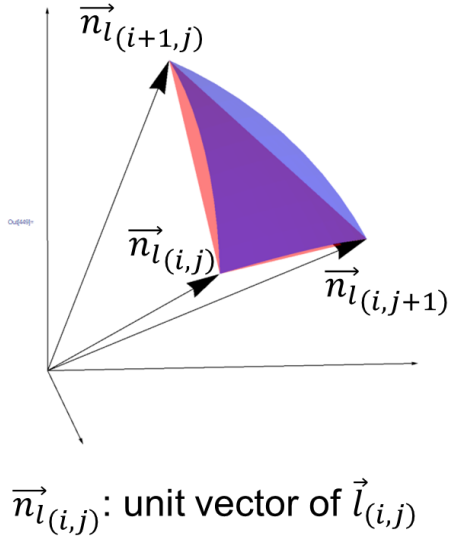

Fig. S3. Schematic drawing of the solid angle, i.e. the area of curved surface colored in blue, calculated by the L'Huilier's formula. The red triangle is the area calculated using the formular (S1). The area of the flat triangle colored in red is smaller than that of the curved surface in blue, providing the smaller skyrmion number of 0.65.

### 4. Full expression of interaction Hamiltonian

First, let us start to derive the ferromagnetic interaction with a misaligned piezoelectric axis. Figure S4 shows the displacement distribution of *x*- and *y*-aligned linearly polarized modes in two coupled membrane resonators with different couplings $g_1$ and $g_2$.

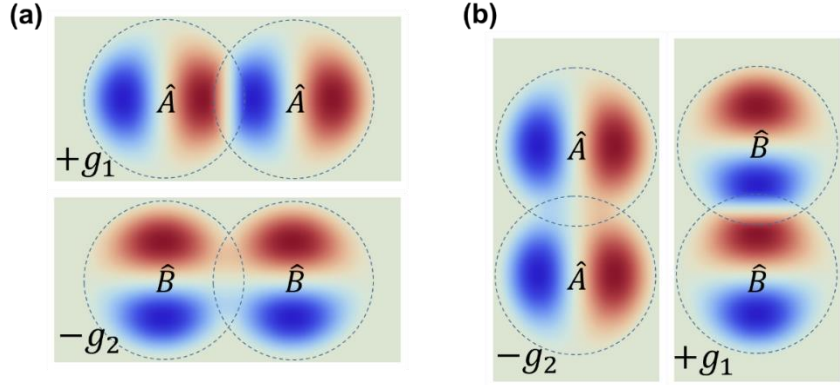

Figure S4. Schematic drawing of the coupling of two modes $\hat{A}$ and $\hat{B}$, which are defined as the (a) x- and (b) y-aligned linearly polarized modes.

Denoting the tilt angle as $\rho$, the displacements of the two modes $A$ and $B$ aligned to two piezoelectric axes are

$$q_{1A} = \hat{q}_{1A} \cos\rho - \hat{q}_{1B} \sin\rho, \quad q_{1B} = \hat{q}_{1A} \sin\rho + \hat{q}_{1B} \cos\rho$$
$$q_{2A} = \hat{q}_{2A} \cos\rho - \hat{q}_{2B} \sin\rho, \quad q_{2B} = \hat{q}_{2A} \sin\rho + \hat{q}_{2B} \cos\rho$$

From Fig. S4, the interaction Hamiltonian is given by

| x-axis | $h_{EX-x} = g_1 \hat{q}_{1A} \hat{q}_{2A} - g_2 \hat{q}_{1B} \hat{q}_{2B}$ | $g_1 > 0, g_2 > 0$ |
|---|---|---|
| y-axis: | $h_{EX-y} = -g_2 \hat{q}_{1A} \hat{q}_{2A} + g_1 \hat{q}_{1B} \hat{q}_{2B}$ | |

Table S2. Interaction Hamiltonian of two membrane resonators.

By defining $g_S = (g_1 + g_2)/2$, $g_{AS} = (g_1 - g_2)/2$, we get

$$h_{EX-x} = g_S(\hat{q}_{1A}\hat{q}_{2A} - \hat{q}_{1B}\hat{q}_{2B}) + g_{AS}[\hat{q}_{1A}\hat{q}_{2A} + \hat{q}_{1B}\hat{q}_{2B}]$$
$$h_{EX-y} = -g_S(\hat{q}_{1A}\hat{q}_{2A} - \hat{q}_{1B}\hat{q}_{2B}) + g_{AS}[\hat{q}_{1A}\hat{q}_{2A} + \hat{q}_{1B}\hat{q}_{2B}]$$

To show that the ferromagnetic terms are dominant, we redefine $\hat{q}_{2A}$ (and those on all the even sites) as $-\hat{q}_{2A}$ in x-axis, and $\hat{q}_{2B}$ (and those on all the odd sites) as $-\hat{q}_{2B}$ in y-axis. The interaction Hamiltonian and the transformation becomes

$$h_{EX-x} = -g_S(\hat{q}_{1A}\hat{q}_{2A} + \hat{q}_{1B}\hat{q}_{2B}) - g_{AS}[\hat{q}_{1A}\hat{q}_{2A} - \hat{q}_{1B}\hat{q}_{2B}]$$
$$h_{EX-y} = -g_S(\hat{q}_{1A}\hat{q}_{2A} + \hat{q}_{1B}\hat{q}_{2B}) + g_{AS}[\hat{q}_{1A}\hat{q}_{2A} - \hat{q}_{1B}\hat{q}_{2B}]$$
$$q_{1A} = \hat{q}_{1A} \cos\rho - \hat{q}_{1B} \sin\rho, \quad q_{1B} = \hat{q}_{1A} \sin\rho + \hat{q}_{1B} \cos\rho$$
$$q_{2A} = \hat{q}_{2A} \cos\rho + \hat{q}_{2B} \sin\rho, \quad q_{2B} = -\hat{q}_{2A} \sin\rho + \hat{q}_{2B} \cos\rho,$$

from which we obtain

$$h_{EX-x} = -g_S \cos 2\rho \, (q_{1A}q_{2A} + q_{1B}q_{2B}) - g_{AS}(q_{1A}q_{2A} - q_{1B}q_{2B})$$
$$\qquad - g_S \sin 2\rho \, (q_{1B}q_{2A} - q_{1A}q_{2B})$$
$$h_{EX-y} = -g_S \cos 2\rho \, (q_{1A}q_{2A} + q_{1B}q_{2B}) + g_{AS}(q_{1A}q_{2A} - q_{1B}q_{2B})$$

$$-g_S \sin 2\rho \, (q_{1B}q_{2A} - q_{1A}q_{2B})$$

By applying rotating frame approximation, we get

$$h_{EX-x} = -\frac{g_S}{2}[\cos 2\rho \, (c_{1A}c_{2A} + s_{1A}s_{2A} + c_{1B}c_{2B} + s_{1B}s_{2B})$$
$$+ \sin 2\rho \, (c_{1B}c_{2A} + s_{1B}s_{2A} - c_{1A}c_{2B} - s_{1A}s_{2B})]$$
$$-\frac{g_{AS}}{2}(c_{1A}c_{2A} + s_{1A}s_{2A} - c_{1B}c_{2B} - s_{1B}s_{2B})$$

$$h_{EX-y} = -\frac{g_S}{2}[\cos 2\rho \, (c_{1A}c_{2A} + s_{1A}s_{2A} + c_{1B}c_{2B} + s_{1B}s_{2B})$$
$$+ \sin 2\rho \, (c_{1B}c_{2A} + s_{1B}s_{2A} - c_{1A}c_{2B} - s_{1A}s_{2B})]$$
$$+\frac{g_{AS}}{2}(c_{1A}c_{2A} + s_{1A}s_{2A} - c_{1B}c_{2B} - s_{1B}s_{2B})$$

(S2)

The second terms in brackets give the expression (10) in main text. Next, we apply the parametric coupling modulation at $\omega_0$,

$$g_1 = g_{10} + g_{1p} \cos 2\omega_0 t, \quad g_2 = g_{20} + g_{2p} \cos 2\omega_0 t$$
$$g_{S0} = (g_{10} + g_{20})/2, \quad g_{AS0} = (g_{10} - g_{20})/2$$
$$g_{Sp} = (g_{1p} + g_{2p})/2, \quad g_{ASp} = (g_{1p} - g_{2p})/2$$

Then, the interaction can be divided into static and parametric terms,

$$h_{EX-x} = h_{EX0-x} + h_{EXp-x}$$
$$h_{EX-y} = h_{EX0-y} + h_{EXp-y}$$

The first terms are identical to (**S2**). The second terms, i.e. the parametric coupling, are

$$h_{EXp-x} = -\frac{g_{Sp}}{4}[\cos 2\rho \, (c_{1A}c_{2A} - s_{1A}s_{2A} + c_{1B}c_{2B} - s_{1B}s_{2B})$$
$$+ \sin 2\rho \, (c_{1B}c_{2A} - s_{1B}s_{2A} - c_{1A}c_{2B} + s_{1A}s_{2B})]$$
$$-\frac{g_{ASp}}{4}(c_{1A}c_{2A} - s_{1A}s_{2A} - c_{1B}c_{2B} + s_{1B}s_{2B})$$

$$h_{EXp-y} = -\frac{g_{Sp}}{4}[\cos 2\rho \, (c_{1A}c_{2A} - s_{1A}s_{2A} + c_{1B}c_{2B} - s_{1B}s_{2B})$$
$$+ \sin 2\rho \, (c_{1B}c_{2A} - s_{1B}s_{2A} - c_{1A}c_{2B} + s_{1A}s_{2B})]$$
$$+\frac{g_{ASp}}{4}(c_{1A}c_{2A} - s_{1A}s_{2A} - c_{1B}c_{2B} + s_{1B}s_{2B})$$

(S3)

The second terms in brackets give the expression (11) in main text by replacing $g_{Sp} \to g_{pc}$. We used these full expressions of (**S2**) and (**S3**) in the numerical simulation shown in Fig. 6.

## 5. Neel-type DM interactions

To generate a Neel-type skyrmion, $\vec{n}_{12} = (0,1,0)$ for a resonator pair aligned to the x-axis, and $h_{DM} = g_{DM}(m_1^z m_2^x - m_1^x m_2^z)$ and $\vec{n}_{12} = (-1,0,0)$ for a resonator pair aligned to the y-axis, $h_{DM} = g_{DM}(-m_1^y m_2^z + m_1^z m_2^y)$. If we replace $\vec{m}$ by $\vec{l}$, we obtain the quadrature DM interaction from as follows.

$$h_{DM-x}^{(L)} = \frac{g_{DM}}{\sqrt{2}}[s_{1A}c_{2A} - c_{1A}s_{2A} + c_{1B}c_{2A} - c_{1A}c_{2B}],$$

$$h_{DM-y}^{(L)} = -\frac{g_{DM}}{\sqrt{2}}[s_{1B}s_{2A} - s_{1A}s_{2B} + s_{1B}c_{2B} - c_{1B}s_{2B}].$$

Similarly, we can define the DM interaction for $\vec{r}$ as

$$h_{DM-x}^{(R)} = \frac{g_{DM}}{\sqrt{2}}[s_{1A}c_{2A} - c_{1A}s_{2A} - c_{1B}c_{2A} + c_{1A}c_{2B}]$$

$$h_{DM-y}^{(R)} = -\frac{g_{DM}}{\sqrt{2}}[s_{1B}s_{2A} - s_{1A}s_{2B} - s_{1B}c_{2B} + c_{1B}s_{2B}]$$

We can eliminate the sine and cosine mixing terms by making a linear combination,

$$h_{DM-x}^{(Neel)} = \frac{h_{DM-x}^{(R)} - h_{DM-x}^{(L)}}{\sqrt{2}} = -g_{DM}[c_{1B}c_{2A} - c_{1A}c_{2B}]$$

$$h_{DM-y}^{(Neel)} = \frac{h_{DM-y}^{(R)} + h_{DM-y}^{(L)}}{\sqrt{2}} = -g_{DM}[s_{1B}s_{2A} - s_{1A}s_{2B}]$$

In contrast to (8), which creates a Bloch-type skyrmion, these interaction Hamiltonians create the Neel-type skyrmion for $R \pm$ as shown in Fig. 3(c).

## 6. Piezoelectric resonant frequency modulation calculated using a finite element method

To obtain eq. (8), we assumed that the parametric frequency shift has opposite signs between two modes. To confirm the feasibility of this assumption, we calculated the flexural mode frequency as a function of d.c. gate voltage for a single membrane resonator using a finite element method (FEM). The calculation is performed using COMSOL Multiphysics® by discretizing the full three-dimensional geometry of the single and coupled membranes with 55460 tetrahedral elements with the minimum size of 6 nanometers. A second-order partial differential equation was applied with fixed constraints at the circular edge of the membrane(s). As a standard measure to evaluate the quality of meshes, the skewness is calculated giving the average value of 0.845, which is large enough to support the mesh quality. Fig. S5 shows the results, which confirm that the voltage-induced frequency shift has an opposite sign between the piezoelectric axes (mode B and C). We can also confirm that a frequency shift of about 1.5% is obtained at $V_g = 1V$, which is large enough to induce the required parametric excitation of $\Gamma = 0.01$.

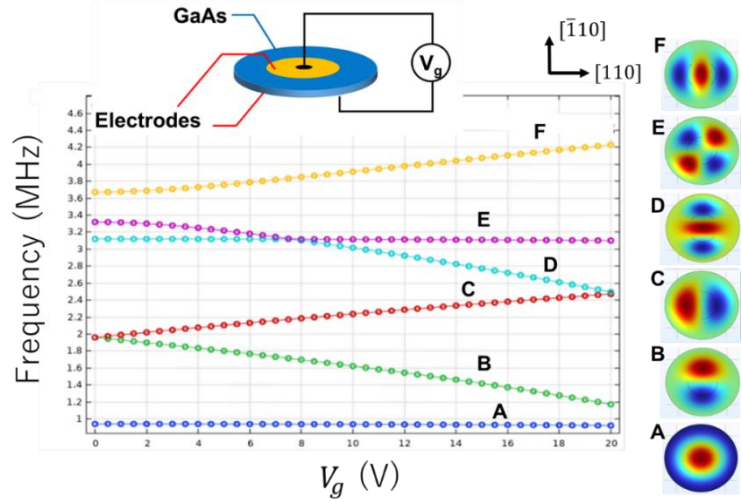

Fig. S5. Calculated flexural mode frequency for a circular membrane GaAs resonator as a function of gate voltage applied between the top and bottom electrodes. The top electrode and membrane radius are 6 $\mu$m and 15 $\mu$m, respectively, and the bottom electrode covers the back surface of the membrane. We set fixed boundary conditions along the edge of the membrane.

7. Stability against the resonance frequency fluctuation

We discuss the robustness of skyrmions against external fluctuations. Fabrication inaccuracies affecting the resonance frequency are some of the most common causes of fluctuations. Here, let us introduce a frequency detuning $\delta_{(i,j)} = (\omega_{0(i,j)} - \omega_0)/\omega_0$ for each resonator obeying a Gaussian distribution. We used a model employing the ideal DM interaction with Eq. (8) and the steady-state textures for various fluctuation $\sigma = \sqrt{\overline{\delta_{(i,j)}^2}}$ are calculated (Fig. S6). When $\sigma \leq 0.02$, the skyrmion texture remains stable against the frequency fluctuation. For a larger $\sigma$, skyrmion-shaped excitations were randomly generated, indicating instability. When the fluctuation is larger than $g_s$, ferromagnetic ordering is not maintained. Nonetheless, since fabrication tolerances of 2% are easily obtained, it should be possible to demonstrate acoustic skyrmions in actual devices.

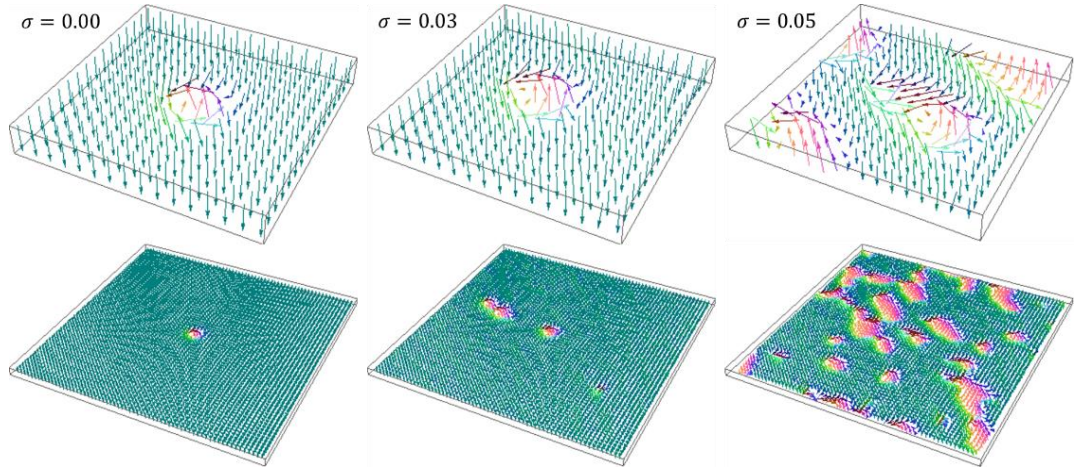

Fig. S6. Calculated textures when frequency fluctuations with different variances are applied to a model with an ideal DMI. $g_s = 0.02$, $\Gamma = 0.01$, $g_{DM} = 0.095$, and damping $1/Q \sim 10^{-4}$.

8. Transition textures

In the parameter region between S-phase and SL-phase, we obtained highly isotropic texture shown in Fig. S7. In this parameter regions, the skyrmion deconfines and occupies the whole lattice in an extremely anisotropic shape.

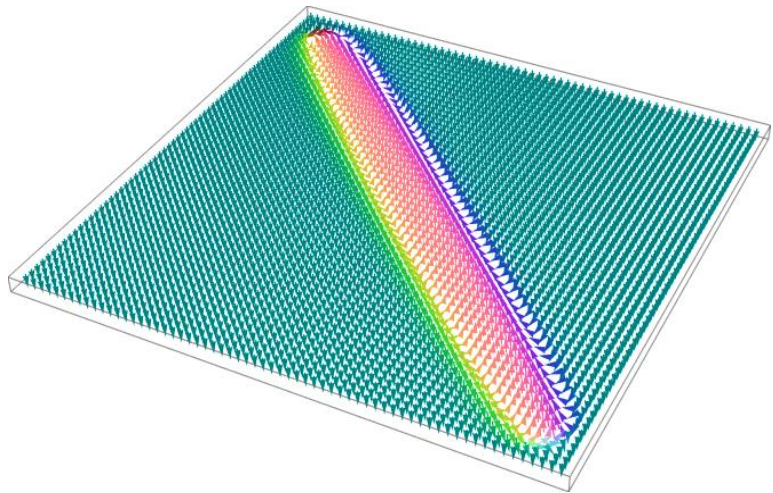

Fig. S7. Calculated textures of $\vec{l}_{(i,j)}$ with an ideal DM interaction. $g_s = 0.02$, $\Gamma = 0.01$, and $g_{DM} = 0.11$. We used the boundary condition fixed to $\vec{l}_{(i,j)} = (0,0,-\sqrt{2}a_0)$ at the periphery.